\documentclass[onecolumn,aps,pra,reprint,superscriptaddress,longbibliography]{revtex4-1}

\usepackage{amsmath}
\usepackage{graphicx}
\usepackage[lofdepth,lotdepth, caption=false]{subfig}
\usepackage{verbatim}
\usepackage{color}
\usepackage{dcolumn}
\usepackage{bm}
\usepackage{miller}
\usepackage{color}
\usepackage{adjustbox}

\usepackage{amsfonts}

\usepackage{setspace}

\usepackage{xr-hyper}
\usepackage{hyperref}

\usepackage[english]{babel}

\newcommand{\beginsupplement}{%
        \setcounter{table}{0}
        \renewcommand{\thetable}{S\arabic{table}}%
        \setcounter{figure}{0}
        \renewcommand{\thefigure}{S\arabic{figure}}%
     }

\begin{document}

\title{Helimagnetism from competing intra- and interchain interactions in CrBr$_2$ and CrI$_2$}

\author{John A.~Schneeloch}
\affiliation{Department of Physics, University of Virginia, Charlottesville,
Virginia 22904, USA}

\author{Matthew B.\ Stone}
\affiliation{Neutron Scattering Division, Oak Ridge National Laboratory, Oak Ridge, Tennessee 37831, USA}

\author{Zachary Morgan}
\affiliation{Neutron Scattering Division, Oak Ridge National Laboratory, Oak Ridge, Tennessee 37831, USA}

\author{Despina Louca}
\thanks{Corresponding author}
\email{louca@virginia.edu}
\affiliation{Department of Physics, University of Virginia, Charlottesville,
Virginia 22904, USA}

\begin{abstract}
CrBr$_2$ and CrI$_2$ are promising platforms for studying the effect of dimensionality on magnetic order, having been isolated in a 3-, 2-, and 1-dimensional form as bulk crystals, monolayers, and individual chains encapsulated in carbon nanotubes. However, the interactions that give rise to their helimagnetic order are unknown. Via inelastic neutron scattering on single crystals, we have determined the exchange interactions of these compounds from the spin wave dispersions, finding that the helimagnetic order arises primarily from competing antiferromagnetic interactions between intrachain and interchain nearest-neighbors. Single-ion anisotropy is substantial, modulating the helical spin rotation and gapping the inelastic intensity at points of branch crossings. As temperature increases, long-range order vanishes but intralayer correlations remain detectable up to at least 50 K. 

\end{abstract}

\maketitle

\pagestyle{plain}

\begin{figure}[t]
\begin{center}
\includegraphics[width=8.6cm]{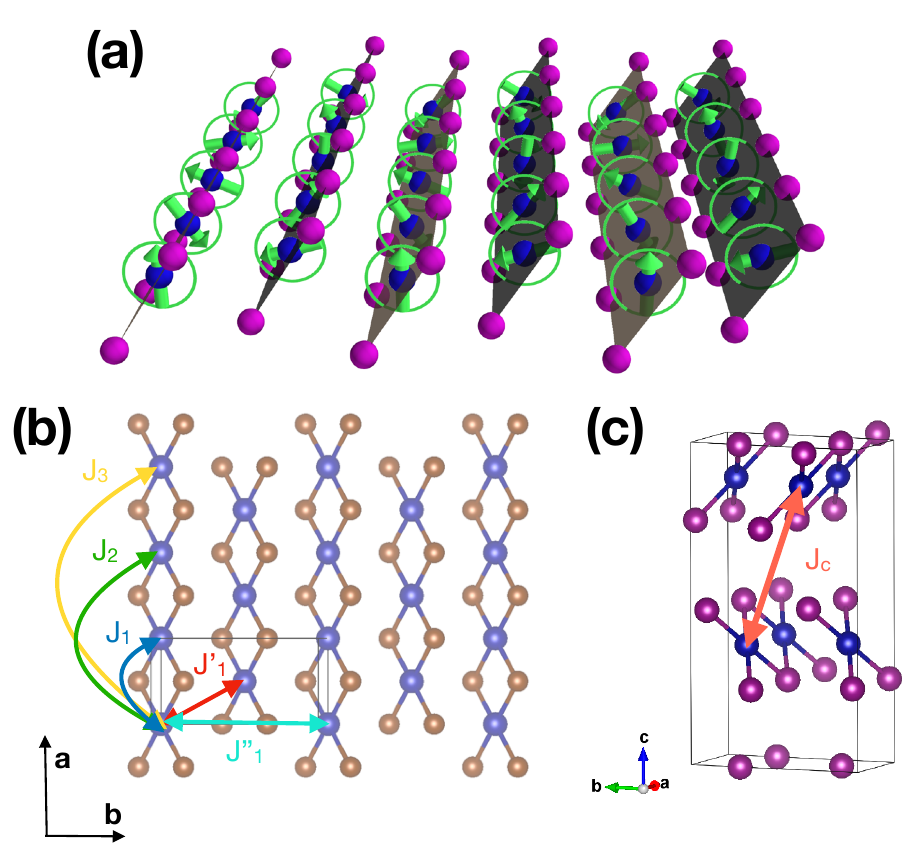}
\end{center}
\caption{(a) A layer of CrBr$_2$, with spins rotating perpendicular to the ribbon-chain axis. (b) Top-down view of a layer of CrBr$_2$ with intralayer exchange bonds labeled. The ribbon chains run vertically. CrI$_2$ is isostructural. (c) Orthorhombic CrI$_2$, the crystal structure of the CrBr$_2$ and CrI$_2$ crystals measured in this study, with the interlayer $J_{c}$ exchange bond labeled. Diagrams (b) and (c) made using \textsc{VESTA} \cite{momma_vesta_2011}.}
\label{fig:Figure1}
\end{figure}

Since the discovery of magnetic order in monolayers of FePS$_3$ \cite{lee_ising-type_2016} and CrI$_3$ \cite{huang_layer-dependent_2017}, the magnetic properties of van der Waals (vdW) layered compounds have received renewed interest \cite{park_2d_2026}. 
One diverse family of vdW-layered compounds are the transition metal dihalides. Each layer has a triangular lattice of cations M$^{2+}$ constructed from edge-sharing MX$_6$ octahedra (X=halogen). Often, these lattices have threefold rotational symmetry, but in CrX$_2$ (X=Br, I) and CuX$_2$ (X=Cl, Br) the lattice is elongated due to the Jahn-Teller (JT) effect, resulting in a ``ribbon chain'' structure (Fig.\ \ref{fig:Figure1}.) 
Layers are weakly coupled to each other via vdW interactions. Thus, the structure of CrBr$_2$ and CrI$_2$ has 1-dimensional (1D) as well as 2-dimensional (2D) aspects. 
(The ribbon-chain motif is also present in CrCl$_2$ \cite{tracy_crystal_1961,winkelmann_structural_1997}, though its chains do not form a vdW-layered structure.) 
The 2D and 1D nature of CrBr$_2$ and CrI$_2$ is highlighted, respectively, by the synthesis of monolayers \cite{li_scanning_2026, karjasilta_molecular_2023, kezilebieke_electronic_2021, cai_molecular_2021,li_single-layer_2020,li_two-dimensional_2023,peng_mott_2020,li_observation_2024, liang_polaron_2025} and the encapsulation of individual ribbon chains within carbon nanotubes \cite{lee_robust_2025}. 
In both compounds, helimagnetic ordering appears below $\sim$17 K \cite{schneeloch_helimagnetism_2024,schneeloch_helimagnetism_2025}. Thus, CrBr$_2$ and CrI$_2$ present an opportunity to study the role of dimensionality on magnetic ordering.

However, the magnetic interactions in CrBr$_2$ and CrI$_2$ are unknown. 
From Curie-Weiss analyses of magnetic susceptibility, effective moments of 4.80 $\mu_B$/Cr and 3.89 $\mu_B$/Cr were found for CrBr$_2$ \cite{schneeloch_helimagnetism_2025} and CrI$_2$ \cite{schneeloch_helimagnetism_2024}, respectively; for CrBr$_2$, the value is close to the ideal 4.90 $\mu_B$ value expected for angular momentum quantum numbers of $L=0$ and $S=2$, while the value for CrI$_2$ is somewhat lower, perhaps reflecting a greater covalency. 
From the chain-like structure, one could speculate that the helical ordering results from competing intrachain interactions. In such a ``$J_1$-$J_2$'' model, with a nearest-neighbor (NN) exchange coupling $J_1$ and a next-nearest-neighbor (NNN) exchange coupling $J_2$, if $J_2$ is sufficiently strongly antiferromagnetic (AFM), the lowest-energy spin structure will be helical   \cite{blundell_magnetism_2001}. CuCl$_2$ and CuBr$_2$, with a similarly Jahn-Teller active Cu$^{2+}$ ion, also have ribbon-chain layered structures and a spin-spiral magnetism (though cycloidal rather than screw-like) \cite{banksMagneticOrderingFrustrated2009,zhaoCuBr2NewMultiferroic2012}; for these and other spin-spiral Cu$^{2+}$ compounds, competing $J_1$ and $J_2$ interactions have been cited as a cause of the magnetic ordering \cite{banksMagneticOrderingFrustrated2009, drechsler_helimagnetism_2007}. Whether this happens to be the case for CrBr$_2$ or CrI$_2$ would require experimental determination of the exchange constants.

Thus, we have performed single-crystal inelastic neutron scattering (INS) measurements on CrBr$_2$ and CrI$_2$ to observe the spin wave dispersion and obtain exchange constants. The primary cause of the spin-spiral ordering is not competition between $J_1$ and $J_2$, but rather between $J_1$ and the NN \emph{interchain} exchange constant $J_1^{\prime}$. 
There is a substantial easy-axis single-ion anisotropy (SIA) that distorts the spin helix and creates avoided crossings in the spin-wave dispersion. 
Interactions are quasi-2D, resulting in intralayer spin correlations being detectable up to at least 50 K, well above the loss of 3D long-range order at $T_N$.

\emph{Elastic neutron scattering} --- We grew crystals of CrBr$_2$ and CrI$_2$ and measured them on the instruments SEQUOIA \cite{stone_comparison_2014} and CORELLI at the Spallation Neutron Source at Oak Ridge National Laboratory. Typically, CrBr$_2$ is monoclinic   \cite{tracy_crystal_1962-1, schneeloch_helimagnetism_2025}, but the crystals grown for these measurements turned out to be in the same orthorhombic polytype as CrI$_2$ (ortho-CrI$_2$) \cite{besrest_structure_1973, schneeloch_helimagnetism_2024}. We have used ortho-CrI$_2$ coordinates throughout this work, i.e., $a \approx 3.65$ \AA, $b \approx 7.10$ \AA, and $c \approx 12.45$ \AA\ for CrBr$_2$; and $a \approx 3.91$ \AA, $b \approx 7.50$ \AA, and $c \approx 13.48$ \AA\ for CrI$_2$. See Supplementary Materials for details on crystal growth, neutron scattering, and the crystal structure \cite{supplement}.

\begin{figure}[t]
\begin{center}
\includegraphics[width=8.6cm]{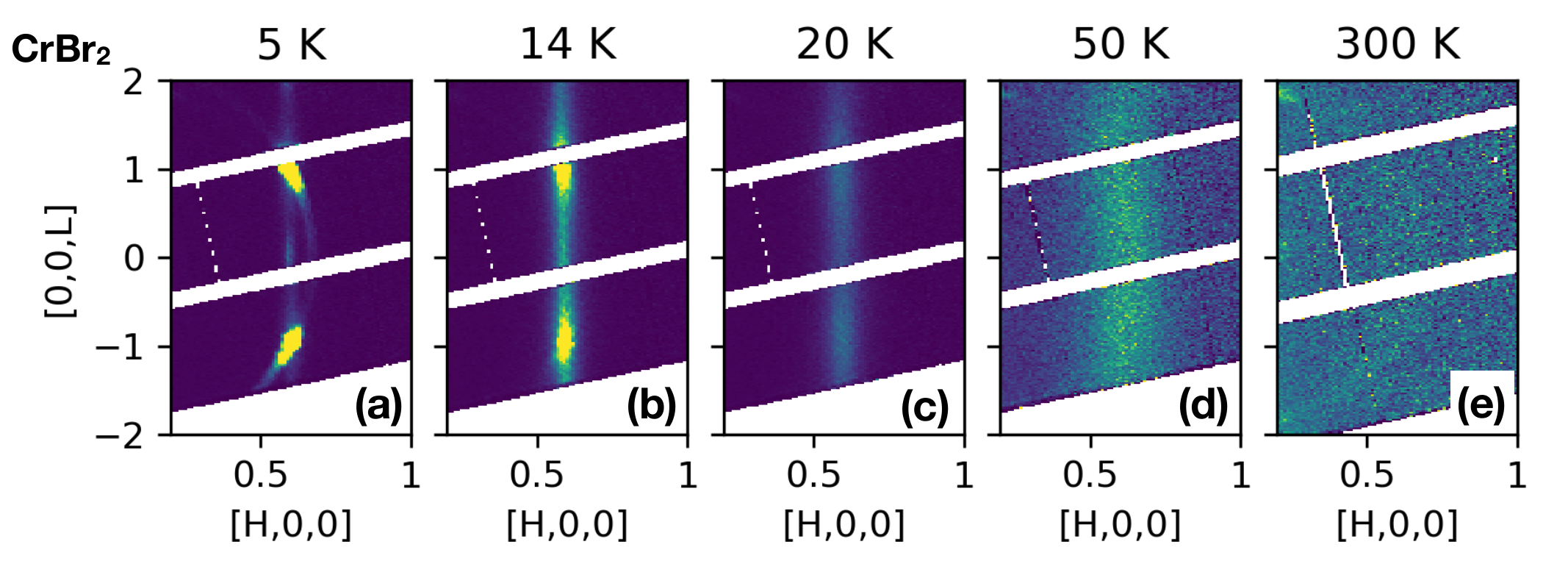}
\end{center}
\caption{Elastic intensity within the H0L plane for CrBr$_2$ at various temperatures. Data were acquired on SEQUOIA at $E_i=17$ meV for (a-d) and $E_i=25$ meV for (e) as described in Sec.\ \ref{sec:ExperimentalDetails} in the Supplemental Materials \cite{supplement}. Intensity scales were the same for 6, 14, and 20 K,  but adjusted to enhance visibility for 50 and 300 K.}
\label{fig:Figure2}
\end{figure}

The magnetic Bragg peaks in Fig.\ \ref{fig:Figure2}(a) show the existence of helimagnetic order in CrBr$_2$ at 5 K. 
These peaks are consistent with the magnetic order reported for CrI$_2$ \cite{schneeloch_helimagnetism_2024}, where peaks are allowed at $(H \pm \delta, K, L)$ for odd-integer $H+K$ and any integer $L$, but with the incommensurability $\delta \approx 0.41$ r.l.u.\ reported for monoclinic CrBr$_2$ \cite{schneeloch_helimagnetism_2025}. 
The small intensity of $(1-\delta,0,0)$ (Figs.\ \ref{fig:Figure1}(a) and \ref{fig:ThreeQPeaks_CrBr2}) suggests that the interlayer phase difference $\alpha \approx 180^{\circ}$ in orthorhombic CrBr$_2$, in contrast to $\alpha \approx 210^{\circ}$ for CrI$_2$ \cite{schneeloch_helimagnetism_2024} (see Sec.\ \ref{sec:MagneticBraggPeakLocations} in the Supplementary Materials for details \cite{supplement}.) Dzyaloshinskii-Moriya (DM) interactions are expected to be larger in CrI$_2$ than CrBr$_2$ due to greater spin-orbit coupling, possibly explaining the larger divergence of $\alpha$ from 180$^{\circ}$ in CrI$_2$. 
On warming, intensity spreads out from the peaks into diffuse rods along $L$, indicating a weakening of interlayer spin coherence. However, intralayer spin correlations remain even for $T>T_N$, up to at least 50 K with only subtle changes in $\delta$. The behavior of CrI$_2$ is similar; see Fig.\ \ref{fig:SuppCrI2H0L} \cite{supplement}. 

A number of indicators point to the presence of a $3 \mathbf{k}_M$ harmonic component in the magnetic ordering of CrBr$_2$ and CrI$_2$.
In CrI$_2$, additional weak peaks are present at $(0.25,0,0)$ and $(0.25,0,\pm1)$ at 5 K, having roughly 0.1\% to 1\% the intensity of the primary magnetic peaks (Fig.\ \ref{fig:ThreeQPeaks} \cite{supplement}.) The weak peak locations are consistent with $(H \pm 3 \delta, K, L)$ ($\delta \approx 0.25$ r.l.u.\ \cite{schneeloch_helimagnetism_2024}) for odd-integer $H+K$. For CrBr$_2$, such peaks were not clearly visible, but as we will explain below, features in the inelastic data of both CrBr$_2$ and CrI$_2$ also point to the presence of a $3 \mathbf{k}_M$ harmonic component. This harmonic can be explained by the inclusion of an easy-axis SIA perpendicular to the ribbon chains with values comparable to CrCl$_2$ (-0.11(2) meV \cite{stone_s2_2013}) and CrCl$_2$(pym) (-0.15(3) meV \cite{pitcairn_low-dimensional_2023}.) 
Calculations show (via energy-minimizing spin structures in \textsc{Sunny} \cite{dahlbom_sunnyjl_2025} using parameters obtained from fitting the inelastic data) that the effect of the SIA is to modulate the helical rotation by twice its primary frequency with amplitudes of $\sim$4$^{\circ}$ for CrBr$_2$ and $\sim$8$^{\circ}$ for CrI$_2$, with the spins remaining in the ($yz$) helical rotation plane. A refinement of the CrI$_2$ magnetic peak intensities to such a model suggested a modulation amplitude of $\sim$6$^{\circ}$, roughly consistent with calculations.

\begin{figure}[t]
\begin{center}
\includegraphics[width=8.6cm]{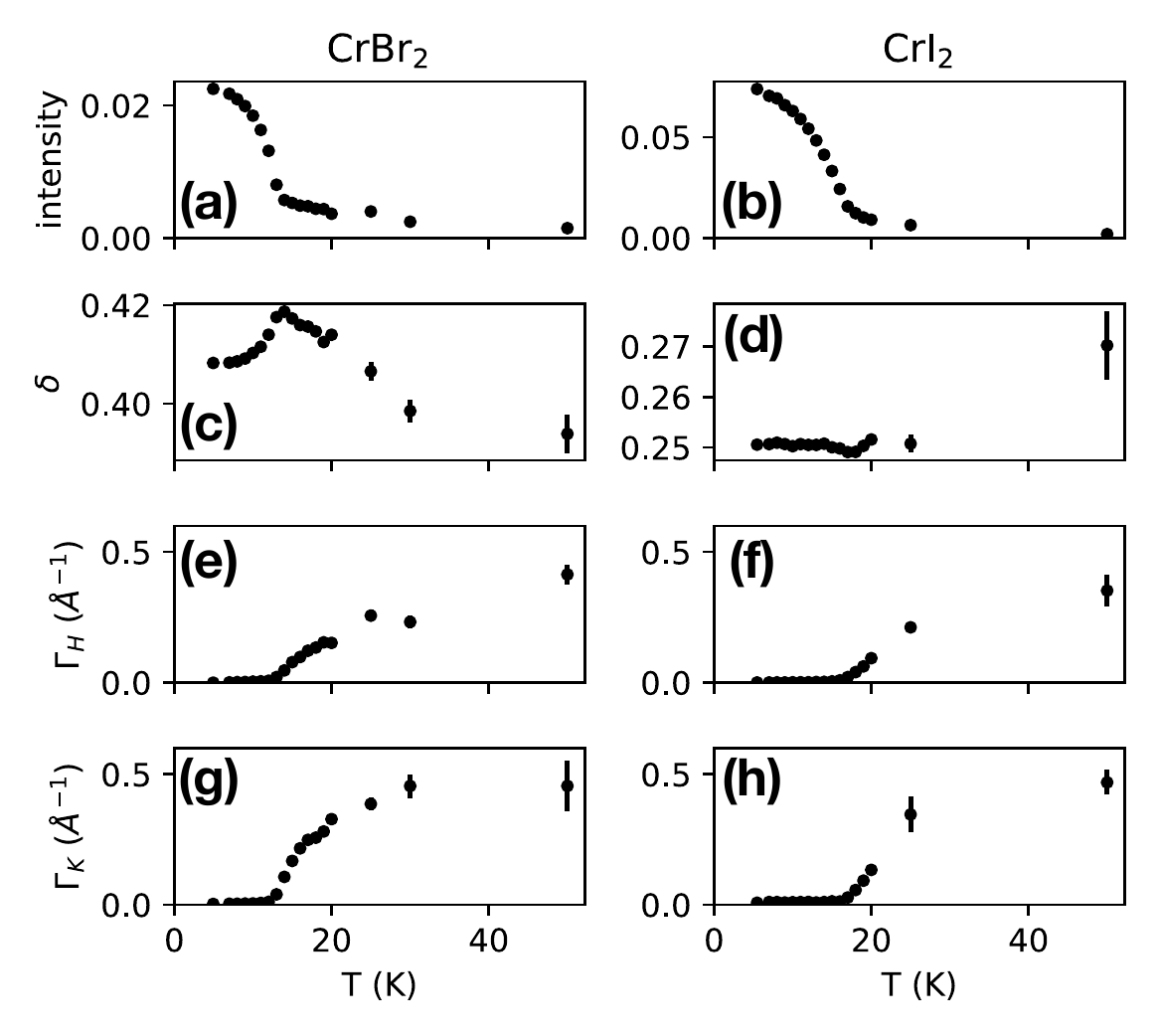}
\end{center}
\caption{Parameters resulting from Voigt-function fits to intensity across the magnetic Bragg peaks (-0.41,-1,0) and (1.25,0,-1) for CrBr$_2$ (a,c,e,g) and CrI$_2$ (b,d,f,h), respectively. Data were obtained on SEQUOIA with $E_i=17$ meV and integrated within 0.1 r.l.u.\ along the $H$ or $K$ directions and 0.2 r.l.u.\ along the $L$ direction. 
The parameters resulted from fits along the $H$ (a-f) or $K$ (g,h) direction. These parameters include the integrated intensity (a,b), the position (c,d), and the Lorentzian FWHM (e-h). The Gaussian component of the Voigt functions was fixed to the low-temperature peak width of each compound's data set, accounting for the instrumental resolution and crystal mosaic.}
\label{fig:Figure3}
\end{figure}

To study changes in ordering with temperature, we fitted Voigt functions to ``cuts'' of intensity vs.\ wavevector along in-plane directions, across $(-0.41, -1 ,0)$ for CrBr$_2$ and $(1.25,0,-1)$ for CrI$_2$. 
The fitted parameters are shown in Fig.\ \ref{fig:Figure3}. 
In CrBr$_2$, the intensity drops as temperature increases until $\sim$14 K, then decreases gradually, likely representing a shift from capturing the peak intensity to that of the diffuse rod. Intensity integrated over all $L$ also showed a decrease (Fig.\ \ref{fig:WideElasticCut}.) 
The incommensurability $\delta$, obtained from the fitted position ($(-\delta,-1,0) \approx (-0.41,-1,0)$), moves anomalously, increasing with temperature until 14 K, then decreasing up to at least 50 K. 
The Lorentzian full-width-at-half-maximum (FWHM) values are plotted as $\Gamma_H$ or $\Gamma_K$ for cuts along $H$ or $K$, indicating how the in-plane correlation lengths $\xi_H = 2/\Gamma_H$ and $\xi_K = 2/\Gamma_K$ change with temperature. 
These correlation lengths appear to become finite above 12 to 13 K, reaching values of 4 to 5 \AA\ at 50 K along both in-plane directions. 
Some of this behavior is consistent with our previous triple-axis spectrometer (TAS) measurements on a monoclinic CrBr$_2$ crystal \cite{schneeloch_helimagnetism_2025}, including the kink in intensity near 14 K and the anomaly in $\delta$. However, the peaks measured via TAS had a total FWHM that was constant with temperature, remaining near $0.03$ \AA$^{-1}$ until the peak vanished beyond 18 K. See Sec.\ \ref{sec:CrBr2TNDiscussion} in the Supplemental Material for a discussion of these contrasting measurements \cite{supplement}.

\begin{table}[t]
\caption{Exchange parameters and single-ion anisotropy (in meV) obtained from fits to data. Positive (negative) values are antiferromagnetic (ferromagnetic.) Uncertainties represent a std.\ dev.\ of statistical error.}
\label{tab:exchangeParameters}
\begin{tabular}{lll|lll}
\hline
& CrBr$_2$ & CrI$_2$ & & CrBr$_2$ & CrI$_2$ \\
$J_1$ & 0.800(6) & 0.407(3) & $J_1^{\prime \prime}$ & & 0.0259(15) \\
$J_2$ & -0.105(6) & 0.0862(18) & $J_c$ & 0.0181(4) & 0.0216(8) \\
$J_3$ & -0.060(4) & & $J_{c,\mathrm{DM}}$ & & 0.0122 \\
$J_1^{\prime}$ & 0.385(4) & 0.465(3) & $D$ & -0.159(4) & -0.1857(15) \\
\hline
\end{tabular}
\end{table}

\begin{figure*}[t]
\begin{center}
\includegraphics[width=18cm]{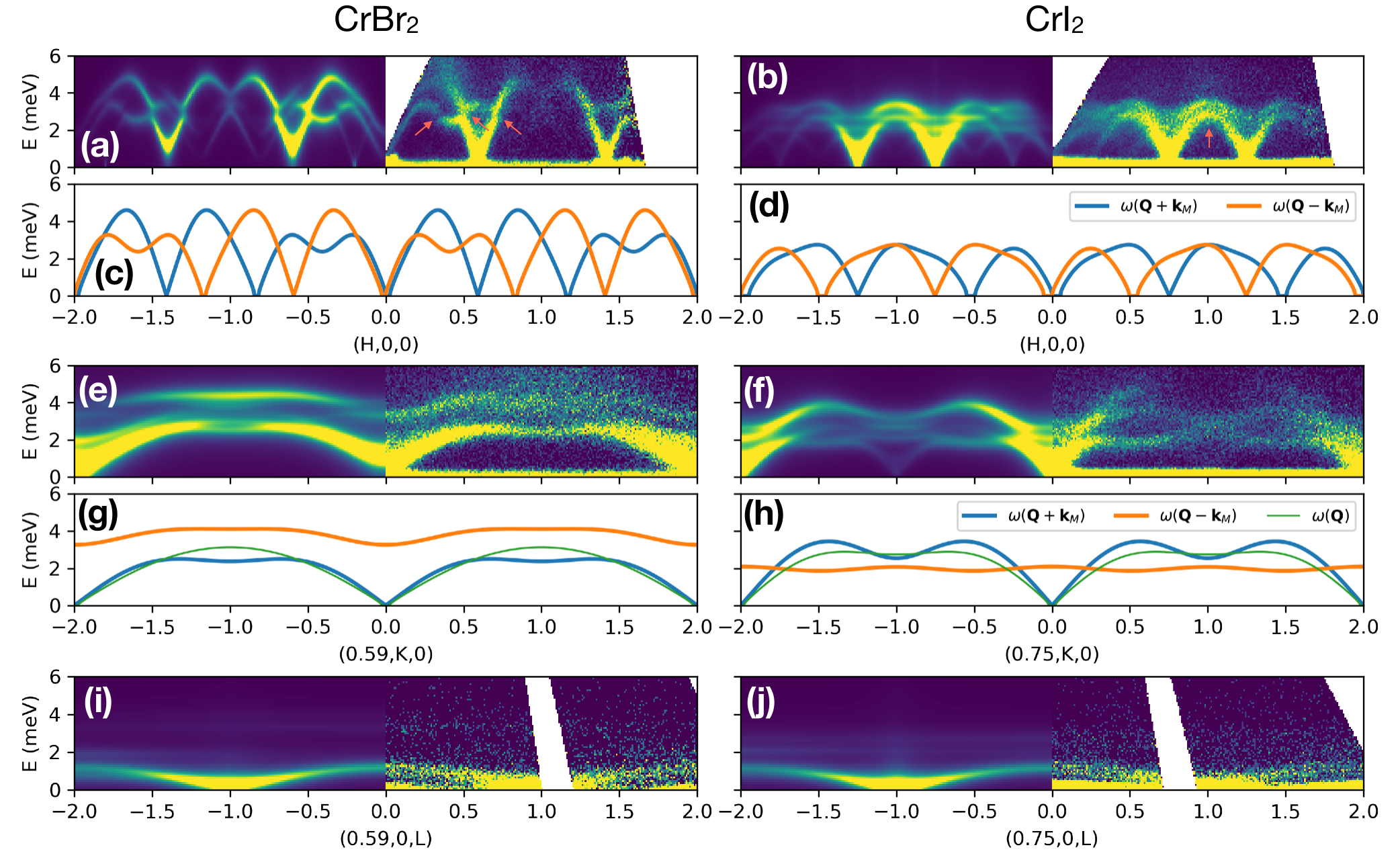}
\end{center}
\caption{Spin wave intensity at 5 K for CrBr$_2$ (a,e,g) and CrI$_2$ (b,f,j), along with the dispersion for simplified single-layer spin-wave models that omit single-ion anisotropy and interlayer coupling. 
Data were taken on SEQUOIA. 
Intensity is plotted as a function of energy transfer $E$, and wavevector along the directions in reciprocal space shown on the horizontal axis, corresponding to the ribbon-chain axis (a,b), the perpendicular intralayer direction (e,f), or the interlayer direction (g,h). The left and right halves of (a,b,e,f,g,h) show, respectively, the calculated and experimental intensity. The arrows in (a) and (b) show avoided crossing points. The integration range was $\pm$0.05 r.l.u.\ along the K direction; $\pm0.02$ r.l.u.\ along the H direction; and, for display purposes, along the entire L range of $\sim$$\pm2$ r.l.u., though the data used for fitting were integrated within $\pm0.2$ r.l.u.}
\label{fig:Figure4}
\end{figure*}

\emph{Spin waves} --- The spin wave dispersion at 5 K is shown in the INS data on the right halves of Figs.\ \ref{fig:Figure4}(a,b,e,f,i,j). The left halves of those subplots show the calculated intensity resulting from the fitted exchange constants and SIA in Table \ref{tab:exchangeParameters}. Fitting was done in \textsc{Sunny} using a $5 \times 1 \times 1$ and $4 \times 1 \times 1$ supercell for CrBr$_2$ and CrI$_2$, respectively; see Sec.\ \ref{sec:fittingInelasticData} in the Supplementary Materials for details \cite{supplement}. The exchange bonds are indicated in Fig.\ \ref{fig:Figure1}. The exchange interactions in the model were isotropic except for the interlayer coupling $J_c$ for CrI$_2$, which had an additional Dzayloshinskii-Moriya (DM) component, $J_{c,\mathrm{DM}}$.

The dominant exchange interactions in CrBr$_2$ and CrI$_2$ are the NN exchange couplings: the intrachain $J_1$ and interchain $J_1^{\prime}$. 
The interlayer coupling is weak; in both CrBr$_2$ and CrI$_2$, $J_c$ is around $0.02$ meV, and the acoustic dispersion along $L$ reaches a maximum of $\sim$1.1 meV. 
Additional intralayer interactions were included to reproduce certain details. For CrBr$_2$, the 3rd-nearest-neighbor intrachain coupling $J_3$ was needed to capture the curvature of the dispersion along $(H,0,0)$, and for CrI$_2$, coupling between next-nearest-neighbor chains via $J_{1}^{\prime \prime}$ was needed to reproduce curvature along $(0.75,K,0)$. In both compounds, $J_2$ is sizable, though apparently ferromagnetic for CrBr$_2$ and AFM for CrI$_2$. 
If we neglect all interactions aside from the two dominant ones, we have a ``$J_1$-$J_1^{\prime}$'' model that can explain the presence of helimagnetic ordering. In this model, the helical angle $\theta$ is minimized for $\theta = 2 \arccos \left(-\frac{J_1^{\prime}}{2 J_1} \right)$. 
Using the fitted exchange constants, we obtain $\theta = 206.7^{\circ}$ for CrBr$_2$ and $\theta = 249.7^{\circ}$ for CrI$_2$, equivalent to 360$^{\circ}$+153.3$^{\circ}$ and 360$^{\circ}$+110.3$^{\circ}$. The angles 153.3$^{\circ}$ and 110.3$^{\circ}$ differ from the experimental values (147.0$^{\circ}$ \cite{schneeloch_helimagnetism_2025} and 89.7$^{\circ}$ \cite{schneeloch_helimagnetism_2024}) by 6.3$^{\circ}$ and 20.6$^{\circ}$ for CrBr$_2$ and CrI$_2$, respectively, showing that the $J_1$-$J_1^{\prime}$ qualitatively explains the helical order for both compounds but agrees better with data for CrBr$_2$.

To explain the features of the data, simplified spin-wave dispersions along the ribbon-chain direction are shown in Figs.\ \ref{fig:Figure4}(c,d,g,h), which were calculated from the fitted exchange constants but with SIA and interlayer coupling omitted. 
The Cr ions on a layer of CrBr$_2$ or CrI$_2$ constitutes a Bravais lattice. A single-$\mathbf{k}_M$ modulated spin structure on a Bravais lattice will generally have three branches visible at a momentum transfer $\mathbf{Q}$ with frequencies $\omega(\mathbf{Q})$, $\omega(\mathbf{Q} + \mathbf{k}_M)$, and $\omega(\mathbf{Q} - \mathbf{k}_M)$ \cite{toth_linear_2015}. Along $(H,0,0)$, however, only the $\omega(\mathbf{Q} + \mathbf{k}_M)$ and $\omega(\mathbf{Q} - \mathbf{k}_M)$ branches are visible, since only spin components perpendicular to $\mathbf{Q}$ contribute to neutron scattering intensity. (For a single-$\mathbf{Q}$ helimagnetic system, using the formalism of Ref.\ \cite{toth_linear_2015}, the tensor $\mathsf{R_2}$ associated with the $\omega(\mathbf{Q})$ branch would be $\mathsf{R_2} = \hat{x} \hat{x}^T$, which would be entirely projected out.) 
Most of the features in the inelastic data can be identified as part of the simplified dispersions. 

Introducing the easy-axis SIA perpendicular to the ribbon chains causes avoided crossings between different dispersion branches. Four such crossing points are indicated by arrows in Fig.\ \ref{fig:Figure4}(a,b), matching corresponding areas in the calculated intensity. 
Notably, the gap near $(0.33,0,0)$ for CrBr$_2$ (Fig.\ \ref{fig:Figure4}(a)) is visible at the crossing between the $\omega(\mathbf{Q}-\mathbf{k}_M)$ and $\omega(\mathbf{Q}-3\mathbf{k}_M)$ branches, even though the $\omega(\mathbf{Q}-3\mathbf{k}_M)$ branch is extremely faint. (See Fig.\ \ref{fig:SuppInelasticBandCrossings} for a plot of the simplified $\omega(\mathbf{Q}\pm3\mathbf{k}_M)$ dispersions. A faint signature of these branches can be seen in Fig.\ \ref{fig:CrBr2FaintDispersion} \cite{supplement}.)  
Overall, the effect of the SIA in CrBr$_2$ and CrI$_2$ is much more subtle than in the collinear antiferromagnets CrCl$_2$ \cite{stone_s2_2013} and CrCl$_2$ \cite{pitcairn_low-dimensional_2023}, where somewhat lower values of the SIA cause gaps of $\sim$2 meV at the bottom of the inelastic intensity.

\emph{Discussion} --- Our results clarify the magnetic interactions in CrBr$_2$ and CrI$_2$. The dominant interactions are the AFM intrachain $J_1$ and the interchain $J_1^{\prime}$, and the interlayer interactions are small, 2-5\% of $J_1$. 
The magnetism of these compounds is thus quasi-2D, with competing intrachain and interchain AFM interactions being the cause of the helical order. 
It is remarkable that $J_1$ is so much smaller in CrI$_2$ than in CrBr$_2$ (0.41 meV vs.\ 0.81 meV), though comparison with the NN exchange constants for CrCl$_2$ (1.13$^{+0.12}_{-0.13}$ meV \cite{stone_s2_2013}) and CrCl$_2$(pym) (1.13(4) meV \cite{pitcairn_low-dimensional_2023}) suggest a trend of decreasing $J_1$ along the series Cl $\rightarrow$ Br $\rightarrow$ I. The interchain coupling $J_1^{\prime}$, in contrast, increases from Br $\rightarrow$ I, perhaps due to increased electron delocalization increasing wavefunction overlap between chains. The other intralayer interactions ($J_2$, $J_3$, and $J_1^{\prime \prime}$) improve the fit by their inclusion but their relevance to the magnetic ordering is unclear. For example, it is not clear if the sign change of $J_2$ between CrBr$_2$ and CrI$_2$ signifies a real change in 2nd-nearest-neighbor intrachain coupling or is due to the limitation of fitting to a model Hamiltonian with exclusively bilinear couplings between spins. 
A fairly large easy-axis SIA is present, -0.16 meV for CrBr$_2$ and -0.19 meV for CrI$_2$, somewhat larger than those found for CrCl$_2$ (-0.11 meV \cite{stone_s2_2013}) and CrCl$_2$(pym) (-0.15 meV \cite{pitcairn_low-dimensional_2023}). 
In contrast, for CrX$_3$ (X=Cl, Br, I), the SIA is very small for CrCl$_3$ (0.014 meV \cite{chen_massless_2021}) and becomes larger for CrBr$_3$ (-0.04 meV \cite{cai_topological_2021}) and CrI$_3$ (-0.123 meV \cite{chen_magnetic_2021}). We suspect that, in CrX$_2$, the JT effect enhances the SIA, while for CrX$_3$, the SIA relies more heavily on the spin-orbit coupling of the anions.

CrBr$_2$ and CrI$_2$ share similarities to some of the other transition metal dihalides \cite{mcguire_crystal_2017}. The vanadium dihalides VCl$_2$ \cite{kadowaki_experimental_1987} and VBr$_2$ \cite{kadowaki_neutron_1985} form a ``120$^{\circ}$'' spin structure, with spins rotating in the $ac$ plane. This structure is equivalent to the spin spiral of CrBr$_2$ or CrI$_2$, except cycloidal rather than screw-like, and with a commensurate helical angle of 120$^{\circ}$. Since VCl$_2$ and VBr$_2$ lack Jahn-Teller distortion, their crystal structures have trigonal symmetry, leading to $J_1 = J_1^{\prime}$ in our notation. This exchange constant is known, from INS, to be AFM in VCl$_2$ \cite{kadowaki_experimental_1987} and VBr$_2$ \cite{kadowaki_neutron_1985}, from which the $J_1$-$J_1^{\prime}$ model implies a 120$^{\circ}$ helical angle. Thus, the incommensurate helical angles of CrBr$_2$ and CrI$_2$ can be seen as a consequence of Jahn-Teller distortion, or, conversely, that VCl$_2$ and VBr$_2$ have spin spiral ordering similar to CrBr$_2$ and CrI$_2$ but for the special case of trigonal symmetry. 

The copper dihalides CuX$_2$ (X=Cl, Br) are especially similar to CrBr$_2$ and CrI$_2$, their layers being isostructural due to the Jahn-Teller distortion of Cu$^{2+}$, and their magnetic order exhibiting a spin spiral structure propagating along ribbon chains \cite{banksMagneticOrderingFrustrated2009,leeInvestigationSpinExchange2012}. Unfortunately, since little is known about their exchange constants, CuX$_2$ serve as examples of compounds where INS measurements of their exchange constants would greatly improve understanding of their magnetic interactions. To our knowledge, no INS studies have been reported on CuCl$_2$. One INS experiment has been reported for CuBr$_2$, but only the very bottom of the spin-wave dispersion was shown in the data \cite{wang_observation_2017}. Thus, CuBr$_2$ has been described as quasi-1D \cite{wang_observation_2017} based on theory, i.e., exchange constants obtained from an energy-mapping method using density functional theory (DFT) calculations \cite{leeInvestigationSpinExchange2012}, but it is difficult to accurately obtain exchange constants in such a manner, as is seen by, e.g., the large discrepancy between the experimental exchange constants in this work and our earlier calculations \cite{schneeloch_helimagnetism_2024}. 
On the other hand, magnetic susceptibility data appear to agree well with a quasi-1D model for CuBr$_2$ \cite{leeInvestigationSpinExchange2012}. If CuCl$_2$ and CuBr$_2$ do, indeed, have quasi-1D magnetic interactions, these compounds would present a contrast with the quasi-2D magnetism of CrBr$_2$ and CrI$_2$ despite similar magnetic ordering. Regardless, it is clear that experimentally obtaining exchange interactions via INS is indispensable to understanding magnetic behavior.

\emph{Conclusion} --- We conducted single-crystal INS and neutron diffraction measurements on CrBr$_2$ and CrI$_2$, obtaining values of their exchange constants and SIA. The dominant exchange interactions are AFM NN intrachain and interchain interactions, whose competition is the cause of the helimagnetic ordering. The interlayer interactions are weak, and intralayer spin correlations are seen to persist well above $T_N$ in both compounds. The SIA distorts the spin spiral, resulting in visible higher-harmonic magnetic Bragg peaks in CrI$_2$ and avoided crossing behavior in the spin-wave dispersions of CrBr$_2$ and CrI$_2$.

\section*{Acknowledgments}

The work at the University of Virginia is supported by the Department of Energy, Grant number DE-FG02-01ER45927. A portion of this research used resources at the Spallation Neutron Source, a DOE Office of Science User Facility operated by Oak Ridge National Laboratory. The beam time was allocated to SEQUOIA on proposal number IPTS-36612, and to CORELLI on proposal number IPTS-36619.


%

\clearpage

\section*{Supplemental Materials}

\beginsupplement

\subsection{Experimental Details}
\label{sec:ExperimentalDetails}
Crystals of CrBr$_2$ and CrI$_2$, were synthesized from the elements as previously reported \cite{schneeloch_helimagnetism_2024,schneeloch_helimagnetism_2025}. After synthesizing precursor crystals in separate ampoules to avoid overpressure, the crystals were combined into a single ampoule, then slowly cooled under a temperature gradient. 

For CrBr$_2$, a horizontal tube furnace was used, with the hot and cold ends cooling, respectively, along 
950 $^{\circ}$C $\rightarrow$ 800 $^{\circ}$C 
and 
900 $^{\circ}$C $\rightarrow$ 750 $^{\circ}$C over 60 hours. Surprisingly, the two crystals measured from this batch had the orthorhombic CrI$_2$ (ortho-CrI$_2$) polytype \cite{besrest_structure_1973} rather than the more typical monoclinic polytype (mono-CrBr$_2$) \cite{tracy_crystal_1962-1,schneeloch_helimagnetism_2025}. 
We are not sure what caused the ortho-CrI$_2$ structure to form; it is possible that the different polytype resulted from slightly higher synthesis temperatures (i.e., 950 $^{\circ}$C at the hot end rather than 900 $^{\circ}$C) or using a controlled temperature gradient rather than a natural temperature gradient of a box or single-zone tube furnace. 
Polytype transition to a monoclinic CrI$_2$ phase (mono-CrI$_2$ \cite{tracy_crystal_1962}) are known to occur in both CrBr$_2$ and CrI$_2$ at elevated temperatures \cite{di_biase_persistent_2026}, but the structural phase diagram is not known in detail. 
For CrI$_2$, a vertical two-zone Bridgman furnace was used, with the upper zone cooling along 950 $^{\circ}$C $\rightarrow$ 800 $^{\circ}$C 
and the lower zone along
900 $^{\circ}$C $\rightarrow$ 750 $^{\circ}$C over 150 hours. CrBr$_2$ and CrI$_2$ are deliquescent, so all handling of these compounds was done under an inert atmosphere. 

Single-crystal inelastic neutron scattering measurements were done at SEQUOIA at the Spallation Neutron Source at Oak Ridge National Laboratory \cite{stone_comparison_2014}. ``High-resolution'' settings were used. The incident neutron energy was $E_i=17$ meV except for data taken at 300 K, for which $E_i=25$ meV. The crystals used for the SEQUOIA experiment weighed 1.4 g for CrBr$_2$ and 3.8 g for CrI$_2$. The crystals were mounted with $\mathbf{c}^*$ oriented upward. Within the plane of the layers, both crystals were single-domain with mosaics of about 1.4(6)$^{\circ}$ for CrBr$_2$ and $2.2(1)^{\circ}$ for CrI$_2$. The crystals were sealed within aluminum cans in a helium glove box prior to measurement. Additional elastic neutron scattering measurements were done on the instrument CORELLI at the same facility.

\subsection{Layer stacking of the CrBr$_2$ and CrI$_2$ crystals}
\label{sec:LayerStacking}

\begin{figure*}[t]
\begin{center}
\includegraphics[width=16cm]{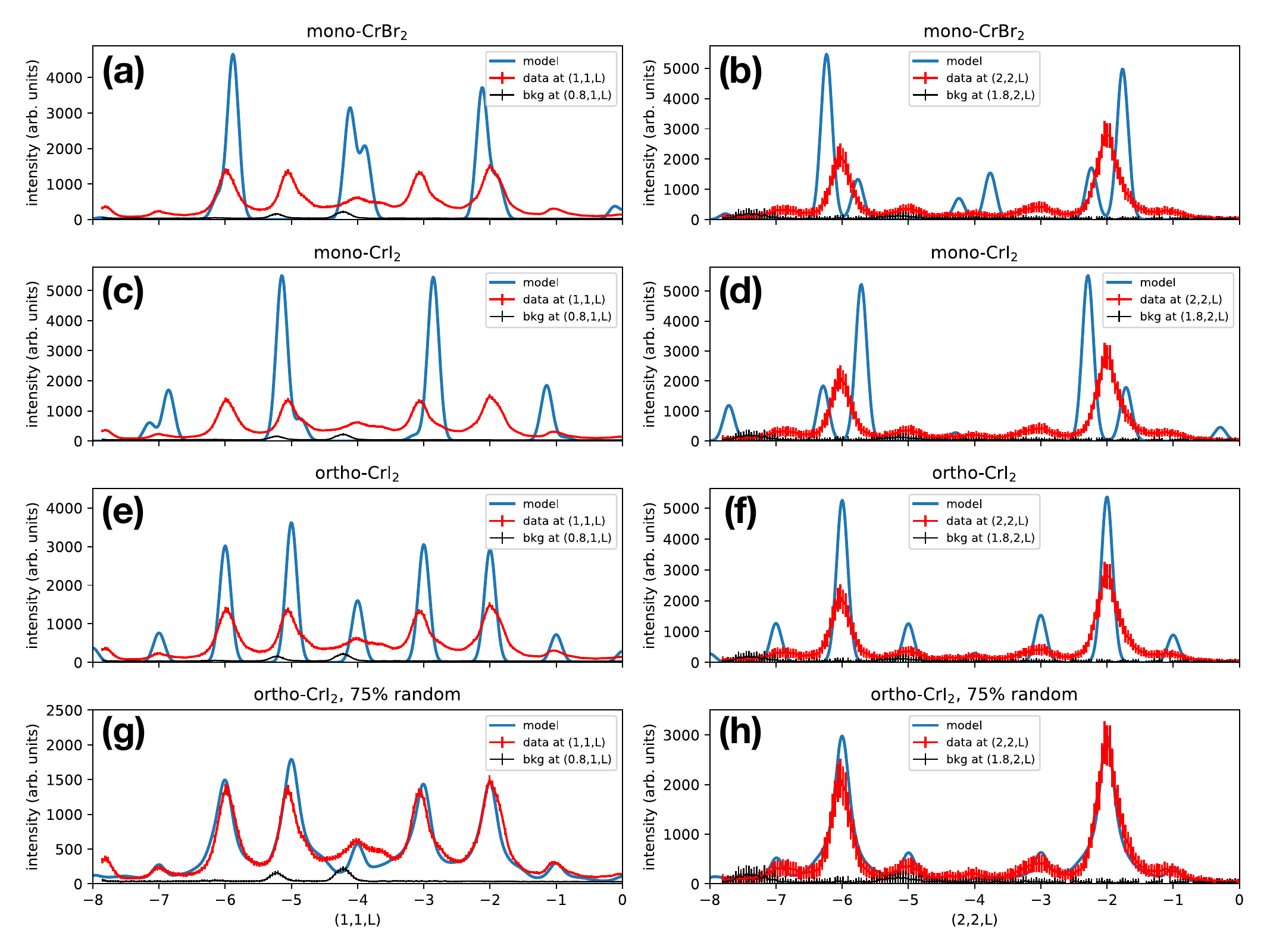}
\end{center}
\caption{Elastic intensity measured with CORELLI on a crystal of CrBr$_2$ at 5 K. The first and second columns show data along $(1,1,L)$ and $(2,2,L)$, respectively, with each row comparing the intensity to a different layer stacking model: monoclinic CrBr$_2$ \cite{tracy_crystal_1962-1} (a,b), monoclinic CrI$_2$ \cite{tracy_crystal_1962} (c,d), orthorhombic CrI$_2$ \cite{besrest_structure_1973} (e,f), and orthorhombic CrI$_2$ with 75\% disordered layer stacking of the kind discussed in Ref.\ \cite{schneeloch_helimagnetism_2024}. For a background comparison, intensity taken at $(0.8,1,L)$ and $(1.8,2,L)$ is also plotted. The data were integrated within $\pm0.05$ r.l.u.\ along the $H$ and $K$ directions.}
\label{fig:CORELLIDiffuse}
\end{figure*}

\begin{figure}[t]
\begin{center}
\includegraphics[width=8cm]{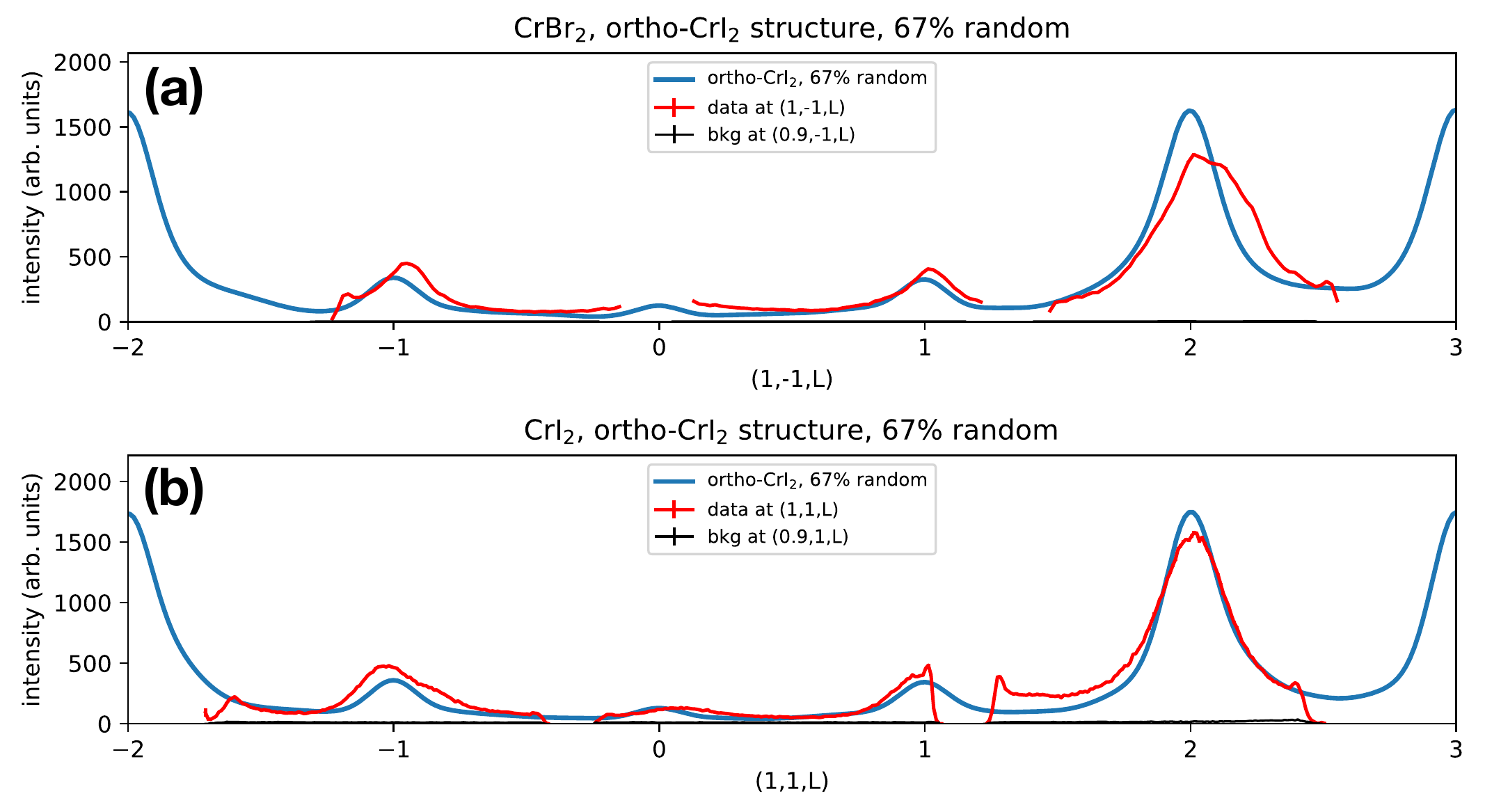}
\end{center}
\caption{Elastic intensity measured with SEQUOIA on (a) CrBr$_2$ along $(1,-1,L)$ and (b) CrI$_2$ along $(1,1,L)$. As a background comparison, data along $(0.9,-1,L)$ or $(0.9,1,L)$ is also shown, though the intensity was very close to zero. Data were integrated within $\pm0.1$ r.l.u.\ along the $H$ and $K$ directions and $-1 \leq E \leq 1$ meV. The data are compared with intensity resulting from a model based on the ortho-CrI$_2$ structure with disordered layer stacking. The gaps in the data are due to gaps between the detectors.}
\label{fig:SEQUOIADiffuse}
\end{figure}

In this section, we compare elastic data taken on CORELLI (Fig.\ \ref{fig:CORELLIDiffuse}) and SEQUOIA (Fig.\ \ref{fig:SEQUOIADiffuse}) with various layer stacking models, finding good agreement with a disordered-stacking ortho-CrI$_2$ structure for both CrBr$_2$ and CrI$_2$. This structure is surprising for CrBr$_2$ given that crystals of CrBr$_2$ are generally found to be monoclinic \cite{tracy_crystal_1962-1,schneeloch_helimagnetism_2025, di_biase_persistent_2026}. Both CrBr$_2$ crystals measured (one on SEQUOIA, and one on CORELLI) were from the same ampoule and thus experienced similar crystal growth conditions to each other, though which crystal growth conditions are responsible for the orthorhombic CrBr$_2$ structure is not clear. 

In Fig.\ \ref{fig:CORELLIDiffuse}, we show intensity of a CrBr$_2$ crystal measured on CORELLI along two lines in reciprocal space, $(1,1,L)$ and $(2,2,L)$. Each row of subplots corresponds to a different model of layer stacking that the data are compared to: 1) the expected monoclinic CrBr$_2$ structure \cite{tracy_crystal_1962-1}; 2) the monoclinic CrI$_2$ structure \cite{tracy_crystal_1962}, which has been seen at high temperature \cite{di_biase_persistent_2026}; 3) the orthorhombic CrI$_2$ structure \cite{besrest_structure_1973}; and 4) the orthorhombic CrI$_2$ structure with 75\% of the volume having random ortho-CrI$_2$-type stacking \cite{schneeloch_helimagnetism_2024}. 

The model intensity was the sum of Gaussian functions with integrated intensity given by the squared structure factor. To model the diffuse scattering, a supercell of $N$ layers was constructed, and the squared structure factor was computed at the supercell Bragg peak locations $(H,K,2L^{\prime}/N)$ for integer $L^{\prime}$. As discussed in Ref.\ \cite{schneeloch_helimagnetism_2024}, each twin of the ortho-CrI$_2$ can be constructed by repeated application of either an ``A'' or ``B'' stacking operation, which are related to each other by translation along the $b$ axis. ``Random ortho-CrI$_2$-type stacking'', thus, means a random choice of A- or B-type stacking at each layer boundary. For Fig.\ \ref{fig:CORELLIDiffuse}(g,h) and Fig.\ \ref{fig:SEQUOIADiffuse}, the diffuse scattering was calculated with 720-layer and 640-layer supercells, respectively. 

In Fig.\ \ref{fig:CORELLIDiffuse}, we see that the peak locations only agree well with the ortho-CrI$_2$ model. This is especially true along $(2,2,L)$, where neither the mono-CrBr$_2$ nor the mono-CrI$_2$ structure would produce the observed peaks at odd-integer $L$. Furthermore, the amount of diffuse scattering observed ($\sim$75\%) is similar to that seen previously in CrI$_2$. For instance, the diffuse scattering in powder CrI$_2$ agreed well with a model where 22\% of the sample had ordered stacking and 78\% had random A/B stacking \cite{schneeloch_helimagnetism_2024}. It is surprising, though, that the amount of random stacking appears to be similar between a ground powder sample and single crystals grown via the Bridgman method. 

Elastic intensity measured on SEQUOIA for both CrBr$_2$ and CrI$_2$ is shown in Fig.\ \ref{fig:SEQUOIADiffuse}. The intensity (along the paths of $(1,-1,L)$ and $(1,1,L)$, which are symmetry-equivalent in the ortho-CrI$_2$ structure) are similar both to each other and to the intensity resulting from the ortho-CrI$_2$ diffuse scattering model. The degree of random stacking present ($\sim$67\%) is similar to that seen on CORELLI.

\subsection{Magnetic Bragg peak locations in reciprocal space}
\label{sec:MagneticBraggPeakLocations}
The intensity of a magnetic Bragg peak is proportional to $\mathbf{F}_M (\mathbf{Q}) \cdot \mathbf{F}^*_M (\mathbf{Q})$, where $\mathbf{F}_M$ is the magnetic structure factor (a vector quantity), $\mathbf{Q}$ is the momentum transfer, and the asterisk denotes the complex conjugate. We let $\mathbf{Q} = \mathbf{G} + \mathbf{k}_M$, where $\mathbf{G}$ is a reciprocal lattice vector and $\mathbf{k}_M$ is the modulation wavevector. For the ortho-CrI$_2$ structure, $\mathbf{k}_M = (1+\delta,0,0)$ in reciprocal lattice units. The formula for the magnetic structure factor of a helical spin spiral is 
\cite{kuindersmaMagneticStructuralInvestigations1981,carpenterElementsSlowNeutronScattering2015} 
\begin{multline}
\label{eq:magEq}
\mathbf{F}_M(\mathbf{G \pm \mathbf{k}_M}) = \frac{\gamma_n r_0}{2 \mu_B} \times \\ \sum_j f(\mathbf{G} \pm \mathbf{k}_M) \frac{|\mathbf{m}_j^{\mathbf{k}_M}|}{2} (\hat{u}_j \pm i \hat{v}_j) \times \\ 
\exp(i (\mathbf{G} \cdot \mathbf{d}_j \mp \phi_j)) 
\exp(-W_j).
\end{multline}
The prefactor $\frac{\gamma_n r_0}{2} = 2.696$ fm. $f(\mathbf{Q})$ is the magnetic form factor. The sum runs over the magnetic ions in the unit cell, which have index $j$ and position $\mathbf{d}_j$. The amplitude of the Fourier component, $|\mathbf{m}_j^{\mathbf{k}_M}|$, is given, in our case, by $g S = 4$, where $g=2$ is the Land\'{e} splitting factor and $S=2$ is the spin atomic number. The vectors $\hat{u}_j$ and $\hat{v}_j$ are perpendicular and define the helical rotation plane. For our case, we let $\hat{u}_j = \hat{y}$ and $\hat{v}_j = \hat{z}$, allowing differences in the helical phases to be captured by the phase factor $\phi_j$. We will neglect the Debye-Waller factor $\exp(-W_j)$. 

We consider a two-layer unit cell, which will have four magnetic atoms, located at $(0,0,0)$, $(0.5,0.5,0)$, $(0,\Delta,0.5)$, and $(0.5, \Delta+0.5,0.5)$. These coordinates describe, up to translation, the locations of the Cr$^{2+}$ ions within the ortho-CrI$_2$ unit cell \cite{besrest_structure_1973}, for which $\Delta \approx 0.672$ r.l.u. These coordinates can also describe the Cr$^{2+}$ locations within unit cells of monoclinic CrBr$_2$ \cite{tracy_crystal_1962-1} or CrI$_2$ \cite{tracy_crystal_1962} (doubled along the $c$ axis and swapping the $x$ and $y$ axes), for which $\Delta=0$. Writing out the structure factor explicitly and simplifying, we have
\begin{multline}
\label{eq:magEq2}
\mathbf{F}_M(\mathbf{G \pm \mathbf{k}}_M) = 
2\gamma_n r_0 
f(\mathbf{G} \pm \mathbf{k}_M) 
(\hat{z} \pm i \hat{y}) \times \\
(1 + e^{i \pi (H+K)})(1 + e^{i \alpha} e^{i 2 \pi K \Delta} e^{i \pi L}
).
\end{multline}
Here, $e^{i \alpha}$ allows for an arbitrary phase between layers. An interlayer phase angle different from $0^{\circ}$ or $180^{\circ}$ would require Dzyaloshinskii-Moriya interactions, which would not be present in the monoclinic structures since the midpoints of the interlayer bonds are centers of inversion symmetry. Thus, the monoclinic phases must have $\alpha = \pi$ or 0, with $\alpha=\pi$ agreeing with the experimentally observed interlayer AFM spin alignment for CrBr$_2$ \cite{schneeloch_helimagnetism_2025}.

From Eq.\ \ref{eq:magEq2}, we can determine whether peaks are allowed at $(H + 1 \pm \delta, K, L)$. 
For the monoclinic CrBr$_2$ or CrI$_2$ unit cells, $\Delta=0$ and $\alpha=\pi$, so peaks are only allowed for odd $L$ in ortho-CrI$_2$ coordinates (or half-integer $L$ in the single-layer coordinates of the monoclinic polytypes.) 
For the ortho-CrI$_2$ structure, if $K=0$, we would expect weak peaks if $\alpha \approx \pi$, and stronger peaks if $\alpha$ diverges from $\pi$ to a greater degree. Thus, $(0.59,0,0)$ being weak in CrBr$_2$ suggests that $\alpha \approx 180^{\circ}$ for  orthorhombic CrBr$_2$, whereas for CrI$_2$, with a relatively strong $(0.75,0,0)$ peak, $\alpha \approx 210^{\circ}$ \cite{schneeloch_helimagnetism_2024}. For $K \neq 0$, though, the $e^{i 2\pi K \Delta}$ factor allows for large magnetic Bragg peaks at $L=0$ to be present.

\subsection{Elastic intensity of CrI$_2$ in $H0L$ plane}

\begin{figure}[t]
\begin{center}
\includegraphics[width=8.6cm]{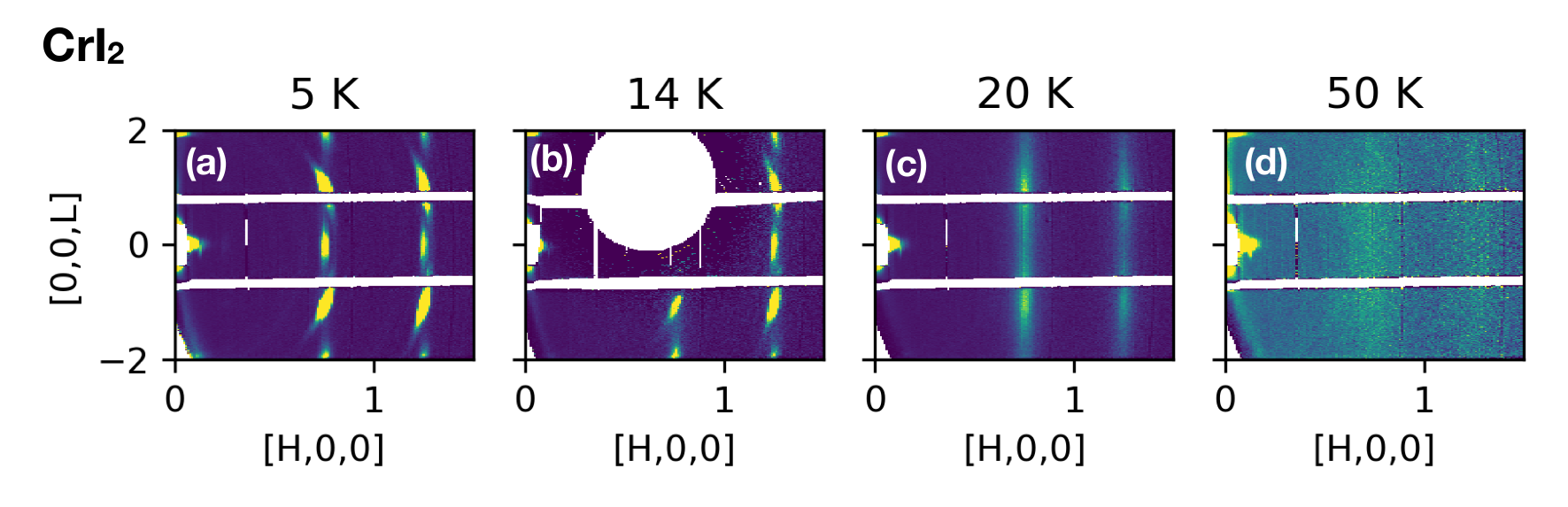}
\end{center}
\caption{Elastic intensity within the H0L plane for CrI$_2$ at various temperatures. Data were taken on SEQUOIA at $E_i=17$ meV. Intensity scales were the same for 6, 14, and 20 K, but adjusted to enhance visibility for 50 K.}
\label{fig:SuppCrI2H0L}
\end{figure}

In Fig.\ \ref{fig:SuppCrI2H0L}, we show the elastic scattering of CrI$_2$ in the $H0L$ plane.

\subsection{Single-ion anisotropy and $3\mathbf{k}_M$-harmonic peaks in CrI$_2$}
\label{sec:elasticSIA}

\begin{figure}[t]
\begin{center}
\includegraphics[width=8cm]{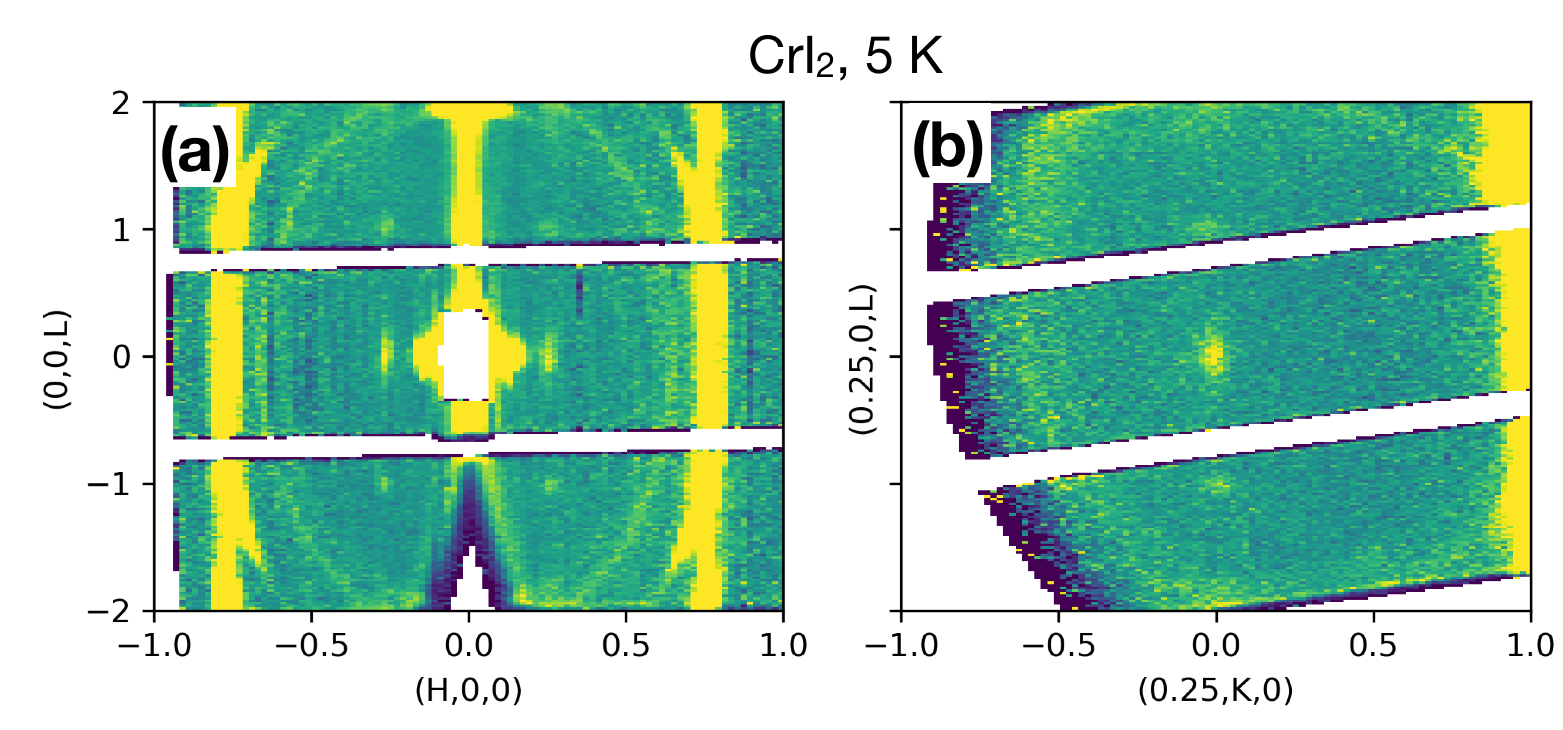}
\end{center}
\caption{Elastic neutron scattering intensity of CrI$_2$ at 5 K, $E_i=17$ meV, displayed in the $(H,0,L)$ plane (a) and the $(0.25,K,L)$ plane (b). Data were acquired on SEQUOIA and integrated within $-1 \leq E \leq 1$ meV and either $-0.1 \leq K \leq 0.1$ r.l.u.\ (a) or $0.2 \leq H \leq 0.3$ r.l.u.\ (b).}
\label{fig:ThreeQPeaks}
\end{figure}

In the main text, we discussed how the magnetic structures of CrI$_2$ and CrBr$_2$ were, predominantly, a single-$\mathbf{k}_M$ helix. For CrI$_2$, with a modulation vector of $\mathbf{k}_M = (1+\delta, 0, 0) \approx (1.25,0,0)$, magnetic Bragg peaks would only be allowed at $(H \pm \delta, K, L)$ for odd $H+K$. However, in Fig.\ \ref{fig:ThreeQPeaks}(a), we see that additional peaks are present at $H \approx 0.25$, $K=0$, and integer $L$, though these are much weaker than the main magnetic Bragg peaks at $(\pm0.75,0,L)$ (not visible in the figure due to saturation of the intensity scale.) Additional weak peaks (not shown) were seen at $(0.75, \pm1, L)$ for integer $L$. The weak peaks are around 2 to 3 orders of magnitude weaker than the main magnetic Bragg peaks. For example, the ratio of the peak at $(0.25,0,1)$ to the relatively strong peak at $(0.75,0,1)$ is  0.0011(4). These weak peaks were absent at 20 K. 
The locations of these weak peaks are consistent with a $3 \mathbf{k}_M$ modulation vector. For example, $(2,0,0) - (1+\delta,0,1) \approx (0.75,0,1)$, and $(4,0,0) - 3 \times(1 - \delta,0,0) \approx (0.25,0,1)$.

\begin{figure}[t]
\begin{center}
\includegraphics[width=8cm]{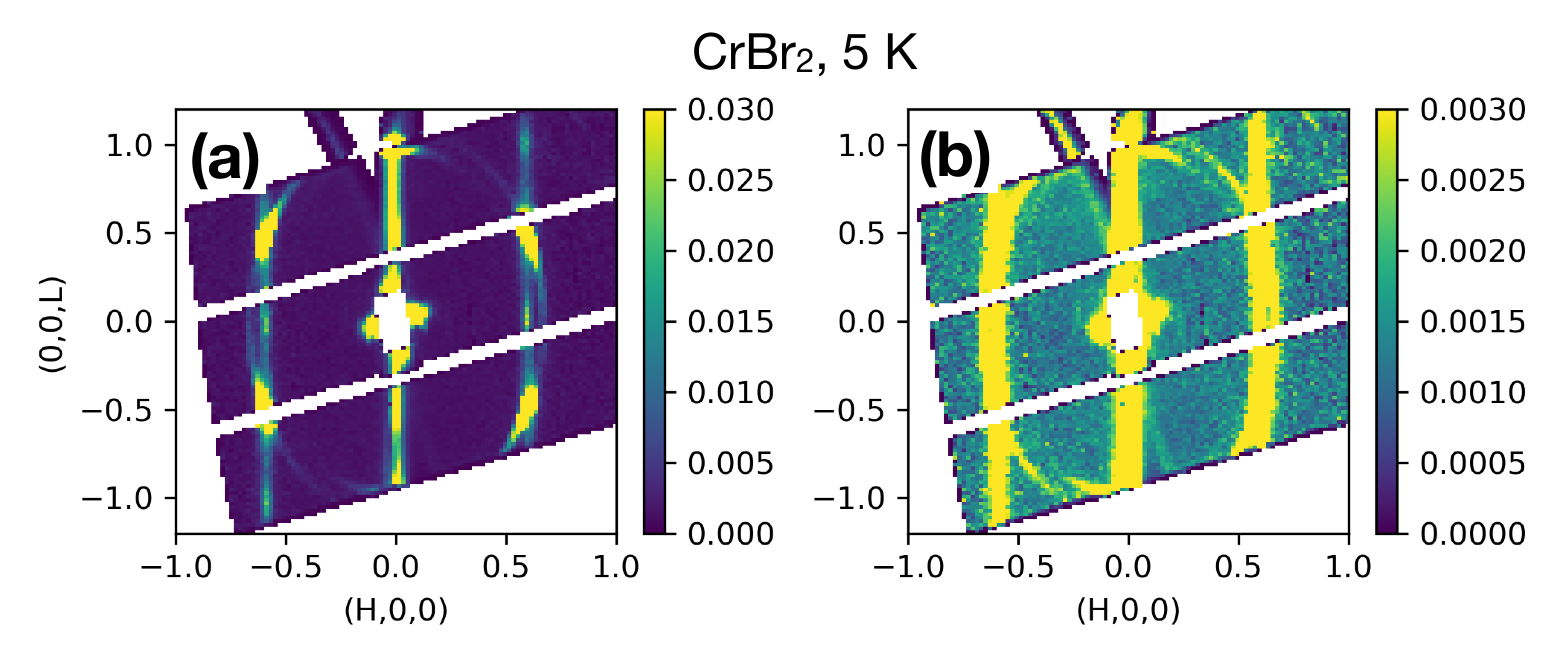}
\end{center}
\caption{(a) Elastic neutron scattering intensity of CrBr$_2$ in the $H0L$ plane at 5 K and $E_i=17$ meV. (b) The same data, but with an enhanced intensity scale. It is ambiguous whether peaks are present at the locations that would be consistent with a $3 \mathbf{k}_M$ harmonic, i.e., $(\pm0.23, 0, L)$ for integer $L$. Data were acquired on SEQUOIA and integrated within $-1 \leq E \leq 1$ meV and $-0.1 \leq K \leq 0.1$ r.l.u.}
\label{fig:ThreeQPeaks_CrBr2}
\end{figure}

\begin{figure}[t]
\begin{center}
\includegraphics[width=8cm]{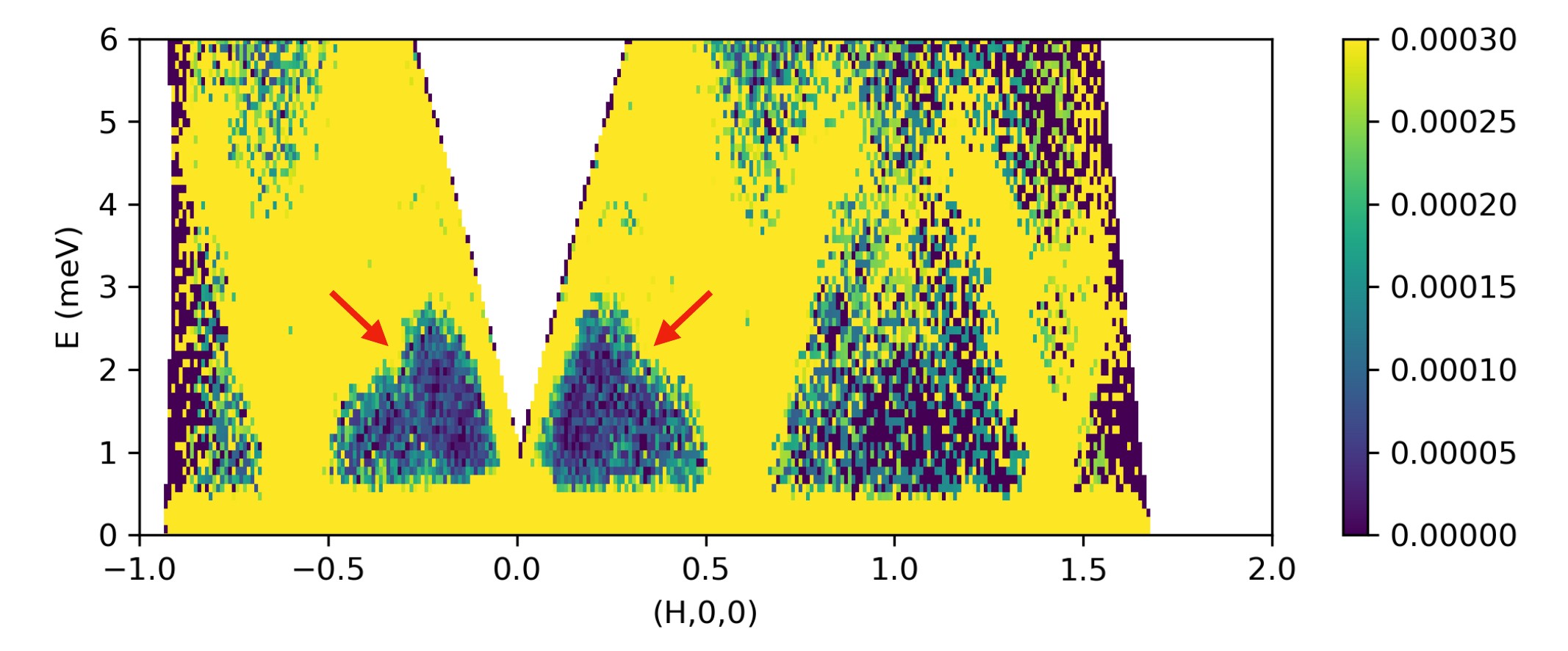}
\end{center}
\caption{Inelastic neutron scattering intensity for CrBr$_2$ at 5 K, taken on SEQUOIA, with intensity integrated over $-0.1 \leq K \leq 0.1$ r.l.u.\ and the entire $L$ range. The intensity scale has been enhanced to highlight the faint third-order harmonic component of the dispersion, plausibly dispersing from the positions indicated by the arrows down to the elastic line at $H \approx \pm 0.23$ r.l.u.}
\label{fig:CrBr2FaintDispersion}
\end{figure}

For CrBr$_2$, the $3 \mathbf{k}_M$ peaks would be located at $(H \pm 3 \delta, K, L) \approx (H \pm 1.23, K, L)$ for odd-integer $H+K$. Thus, within the H0L plane (Fig.\ \ref{fig:ThreeQPeaks_CrBr2}), we would expect $3\mathbf{k}_M$ peaks to be visible at $(\pm 0.23, 0, L)$ for integer $L$. Such peaks are not clearly visible in Fig.\ \ref{fig:ThreeQPeaks_CrBr2}, even with an enhanced intensity scale, but faint signs of a dispersion emanating from these points can be seen in Fig.\ \ref{fig:CrBr2FaintDispersion}. The corresponding feature in CrI$_2$, dispersing from $(0.25,0,0)$ in Fig.\ \ref{fig:Figure4}(b), is much more clear.

\subsection{Integrated magnetic elastic intensity}
\begin{figure}[t]
\begin{center}
\includegraphics[width=8cm]{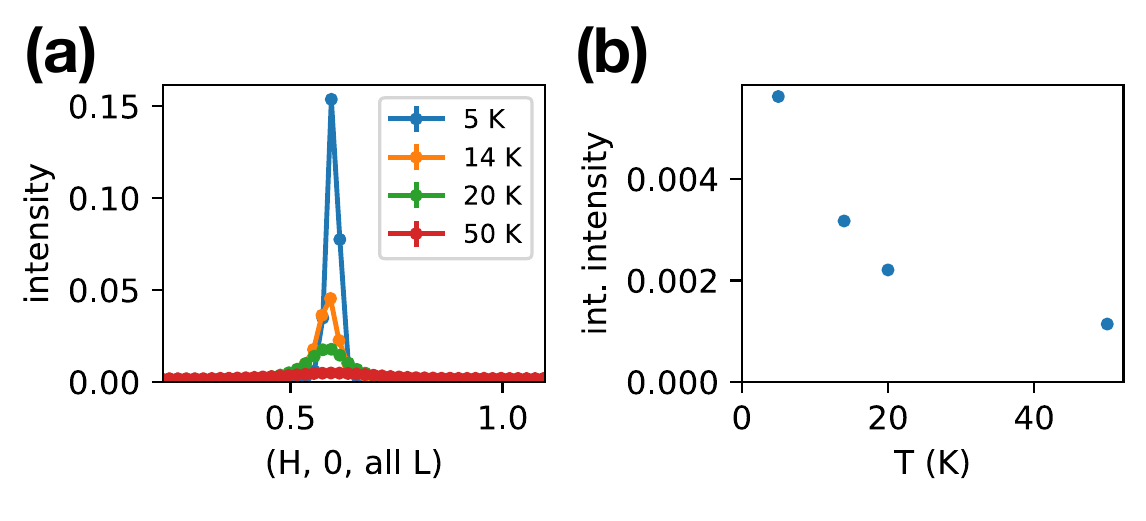}
\end{center}
\caption{(a) Elastic intensity of CrBr$_2$ along $(H,0,0)$, integrated within $-1 \leq E \leq 1$ meV, $-0.3 \leq K \leq 0.3$, and over the full $L$ range in the data. The peak near $H=0.59$ includes contributions from magnetic Bragg peaks or diffuse rods along $(0.59, 0, L)$. (b) Intensity of data in (a) integrated within $0.25 \leq H \leq 1$ after subtracting the uniform background present at 5 K. Data were taken on SEQUOIA with $E_i=17$ meV.}
\label{fig:WideElasticCut}
\end{figure}

To get a sense of the degree to which intralayer spin correlations persist to higher temperatures, in Fig.\ \ref{fig:WideElasticCut}(a) we show the elastic neutron scattering intensity of CrBr$_2$ along $(H,0,0)$, integrated over the entire $L$ range in the data. The peak near $H=0.59$ has contributions from both the magnetic Bragg peaks along $(0.59,0,L)$, present only at low temperature, and the diffuse rod of magnetic intensity that arises on warming (Fig.\ \ref{fig:Figure2}.) The integrated intensity of this peak is shown in Fig.\ \ref{fig:WideElasticCut}(b), suggesting that a substantial degree of intralayer spin correlations persist up to 50 K.

\subsection{Long-range and short-range order in CrBr$_2$ from TAS and SEQUOIA measurements}
\label{sec:CrBr2TNDiscussion}
Here, we discuss apparent disagreement in obtaining $T_N$ in our current data on orthorhombic CrBr$_2$ and our previous data on monoclinic CrBr$_2$ \cite{schneeloch_helimagnetism_2025}, and how these sets of measurements can be partly reconciled, as well as the need for additional measurements to resolve the issue. 
When reciprocal space coordinates are shown that reference the monoclinic CrBr$_2$ structure \cite{tracy_crystal_1962-1}, they will be denoted with a subscript $M$, i.e., $(H,K,L)_M$. 

In Ref.\ \cite{schneeloch_helimagnetism_2025}, we performed triple-axis spectrometer measurements with the instrument HB1A on a single crystal of CrBr$_2$. A scan along $(20L)_M$ confirmed that CrBr$_2$ had the typical monoclinic crystal structure, that only one twin was present, and that diffuse scattering (that would indicate stacking disorder) was minimal. (Thus, both the type of layer stacking and the lack of stacking disorder differ from the crystal measured on SEQUOIA, which was orthorhombic with $\sim$75\% of the sample having random stacking of ortho-CrI$_2$ type, as discussed in Sec.\ \ref{sec:LayerStacking}.) The magnetic order was ascertained via scans along $K$ in $(0,K,L)_M$ for various $L$ values. The peak at $(0,1-\delta,1/2)_M$ was fitted with a Gaussian function, and the intensity and position (via $\delta$) were plotted as a function of temperature. The intensity had two apparent kinks, one near 13 K, and one near 17 K. The incommensurability $\delta$ increased from $\sim$0.41, reaching 0.42 at 15 K, then decreased, being below 0.41 at 18 K. We did not report the FWHM, but it remained near $0.03$ \AA\ within uncertainty up to 18 K, beyond which point the peak was no longer visible. The range of the HB1A scans only covered 0.07 r.l.u.\ $\approx 0.12$ \AA$^{-1}$ along the $K$ direction, and thus these scans may have missed a component from the diffuse rod at $(0,0.59,L)$ once its FWHM increased beyond $\sim$$0.12$ \AA$^{-1}$. 

The SEQUOIA data (Fig.\ \ref{fig:Figure3}) show some similarities with the HB1A measurements. The sharp drop in intensity between 12 and 13 K is seen in both compounds. However, the SEQUOIA data show intensity persisting up to 50 K, while the HB1A data show no peak present beyond 18 K. As we noted, though, the HB1A scans had a short range and may have only picked up the narrow long-range-order peak while missing the broader diffuse rod. 
The $\delta$ anomaly has maxima at similar temperatures, 15 K for the HB1A data and 14 K for the SEQUOIA data, and reaches a similar value of $\sim$0.42 r.l.u.\ after starting from $\sim$0.41 r.l.u.\ at low temperature. The behavior beyond this point is somewhat different, though, dropping to $\delta=0.41$ r.l.u.\ by 18 K in the HB1A data but only crossing this point between 20 and 25 K in the SEQUOIA data. 

Overall, changes in magnetic order manifest similarly in the measurements on HB1A and SEQUOIA, but there are differences, the most significant being the apparent onset of short-range intralayer order above 12 to 13 K in the SEQUOIA data while a narrow peak persists up to 18 K in the HB1A data. Obviously, the intralayer interactions are similar in these crystals, as shown by the similar $\delta$ values. 
The strength of the interlayer interactions should differ; in other vdW-layered magnetic materials such as CrCl$_3$ and CrI$_3$, there is evidence that interlayer coupling strength varying due to the stacking types can shift transition temperatures \cite{schneeloch_antiferromagnetic-ferromagnetic_2024,schneeloch_role_2024}. However, the maximum in the $\delta$ anomaly occurring near the same temperature in both crystals suggests that the interlayer coupling strength may be similar, assuming the anomaly is connected to the loss of interlayer spin coherence. 
It is possible that differences in stacking disorder may affect the transition, but Landau-Lifshitz simulations that we have conducted in \textsc{Sunny} suggest that the diffuse rod emerging from the magnetic Bragg peaks on warming is a behavior that occurs even with perfectly ordered layer stacking. 
Finally, it is possible that the SEQUOIA measurements missed the presence of a narrow long-range-order peak once the diffuse rod intensity overwhelmed it, but it is not clear why long-range order would persist up to $\sim$18 K even as intralayer short-range order increases above 12 to 13 K.
 
Ultimately, additional studies are needed to untangle the nature of the transition in CrBr$_2$, to explain the kinks in intensity vs.\ temperature and the anomaly in $\delta$, and to determine how layer stacking variation affects the transition.

\subsection{Fitting inelastic data}
\label{sec:fittingInelasticData}
\begin{figure*}[t]
\begin{center}
\includegraphics[width=18cm]{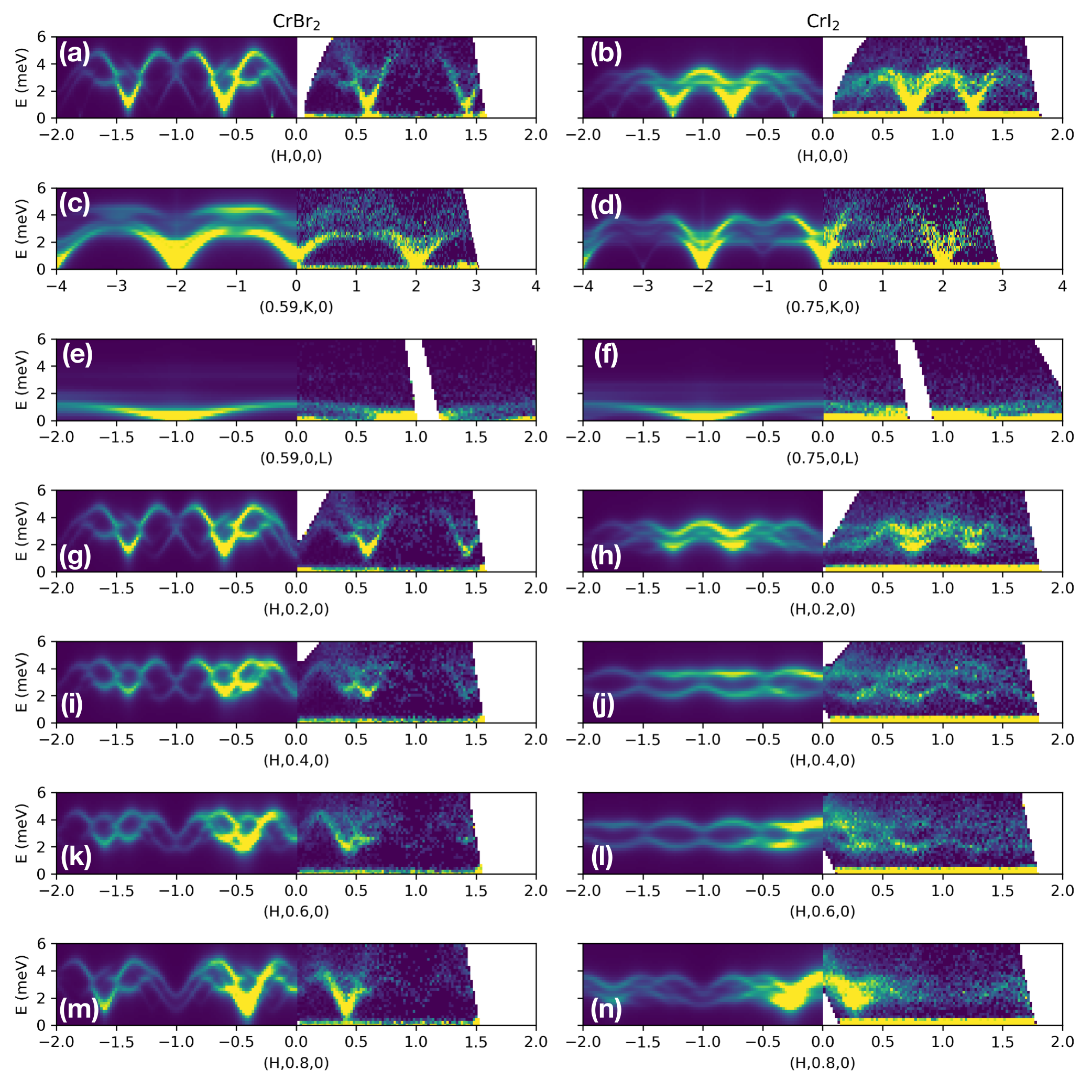}
\end{center}
\caption{INS data used for fitting and obtaining the exchange interactions and single-ion anisotropy. The right half of each subplot shows the data, and the left half shows the calculated intensity. The first and second columns show data for CrBr$_2$ and CrI$_2$, respectively. All data were taken on SEQUOIA at 5 K with $E_i=17$ meV. The data were integrated within $\Delta H = \pm 0.02$ r.l.u., $\Delta K = \pm 0.05$ r.l.u., or 
$\Delta L = \pm 0.2$ r.l.u.\ along perpendicular directions.}
\label{fig:SuppInelastic}
\end{figure*}

Here we present the details of fitting the INS data to find exchange constants and the easy-axis SIA. The data used for fitting are shown in Fig.\ \ref{fig:SuppInelastic}. 
(In contrast to Fig.\ \ref{fig:Figure4}, where much of the data were integrated over all $L$ for visual clarity, in Fig.\ \ref{fig:SuppInelastic} any data selected along a constant-$L$ path were restricted to $L=0$, integrated within a narrow $\Delta L$ window.) 
The fitting was done in \textsc{Sunny} \cite{dahlbom_sunnyjl_2025}. The Land\'{e} factor was set to $g=2$, and the spin/total angular momentum quantum number (with zero orbital angular momentum) was set to $S=2$. The quantity $\chi^2$ was calculated by summing over every pixel in the subplots in Fig.\ \ref{fig:SuppInelastic} for a particular compound, excluding the elastic line $(E < 0.6$ meV.) For CrI$_2$, $E > 4.5$ meV was also excluded to minimize the effect of intensity above the dispersions. 
Linear spin-wave theory was used with a supercell of $5 \times 1 \times 1$ for CrBr$_2$ and $4 \times 1 \times 1$ was used for CrI$_2$. These supercells were chosen so that spins would undergo an approximately whole number of rotations from one end of the supercell to the other. Specifically, for CrBr$_2$, the helical rotation angle is $147.0^{\circ} \approx 360^{\circ} \times \frac{5}{2} = 144^{\circ}$; for CrI$_2$, the helical rotation angle is $89.2^{\circ} \approx 360^{\circ}/4 = 90^{\circ}$. (Although \textsc{Sunny} has a ``spiral'' method for calculating the linear spin-wave theory intensity for an incommensurate structure, this method does not currently allow including SIA, so we used the supercell approach.) The spin structure of the supercell was brought to an energy minimum before each computation of the INS intensity, which allowed the distortion of the helical rotation by the SIA to be incorporated. $\chi^2$ was minimized using the Nelder-Mead algorithm. If a set of parameters resulted in energy instability, a very large value was returned for $\chi^2$ in lieu of a calculated value. 

The exchange constants and SIA are derived from the Heisenberg Hamiltonian
\begin{equation}
    E = \sum_{\langle i,j \rangle_n} \mathbf{S}^{\mathsf{T}}_i \cdot J_n \cdot \mathbf{S}_j + 
    \sum_i (D (\mathbf{S}_i \cdot \hat{p}^{(1)}_i)^2 + 
    D_2 (\mathbf{S}_i \cdot \hat{p}^{(2)}_i)^2).
\end{equation}
Here, $i$ and $j$ are spin site indices. $\langle i,j \rangle_n$ denotes exchange bonds of type $n$, i.e., iterating over the exchange interactions listed in Table \ref{tab:exchangeParameters} and illustrated in Fig.\ \ref{fig:Figure1}. Each bond is counted once in the sum. Most of the exchange interactions are isotropic, but the interlayer interaction is split into an isotropic part $J_c$ and an anisotropic Dzyaloshinskii-Moriya (DM) vector $(J_{c,\mathrm{DM}},0,0)$. For CrI$_2$, the DM contribution was needed to account for an interlayer phase of 210$^{\circ}$ \cite{schneeloch_helimagnetism_2024}; thus, during fitting, the ratio $J_{c,\mathrm{DM}}/J_c$ was fixed to the value needed to reproduce this interlayer phase. For CrBr$_2$, we set $J_{c,\mathrm{DM}} \rightarrow 0$ since the very weak $(0.59,0,0)$ peak implies an interlayer phase very close to 180$^{\circ}$. 
The interlayer exchange constant $J_c$ was fitted separately prior to the final fitting of the other parameters. The coefficient $D$ refers to the primary SIA considered, that is, the easy-axis SIA with $\hat{p}_i^{(1)}$ perpendicular to the ribbon chains. Additionally, a small $(D_2=0.0001$ meV) easy-plane SIA with $\hat{p}_i^{(2)}$ along the ribbon-chain axis was added so that the lowest energy state of the helix would have the helical rotation perpendicular to the ribbon chain axis (i.e., for a screw-like rather than cycloidal spin spiral), as is consistent with previous data \cite{schneeloch_helimagnetism_2024, schneeloch_helimagnetism_2025}.

\subsection{Spin wave branches with 3$\textbf{k}_M$ harmonics included}

\begin{figure*}[t]
\begin{center}
\includegraphics[width=18cm]{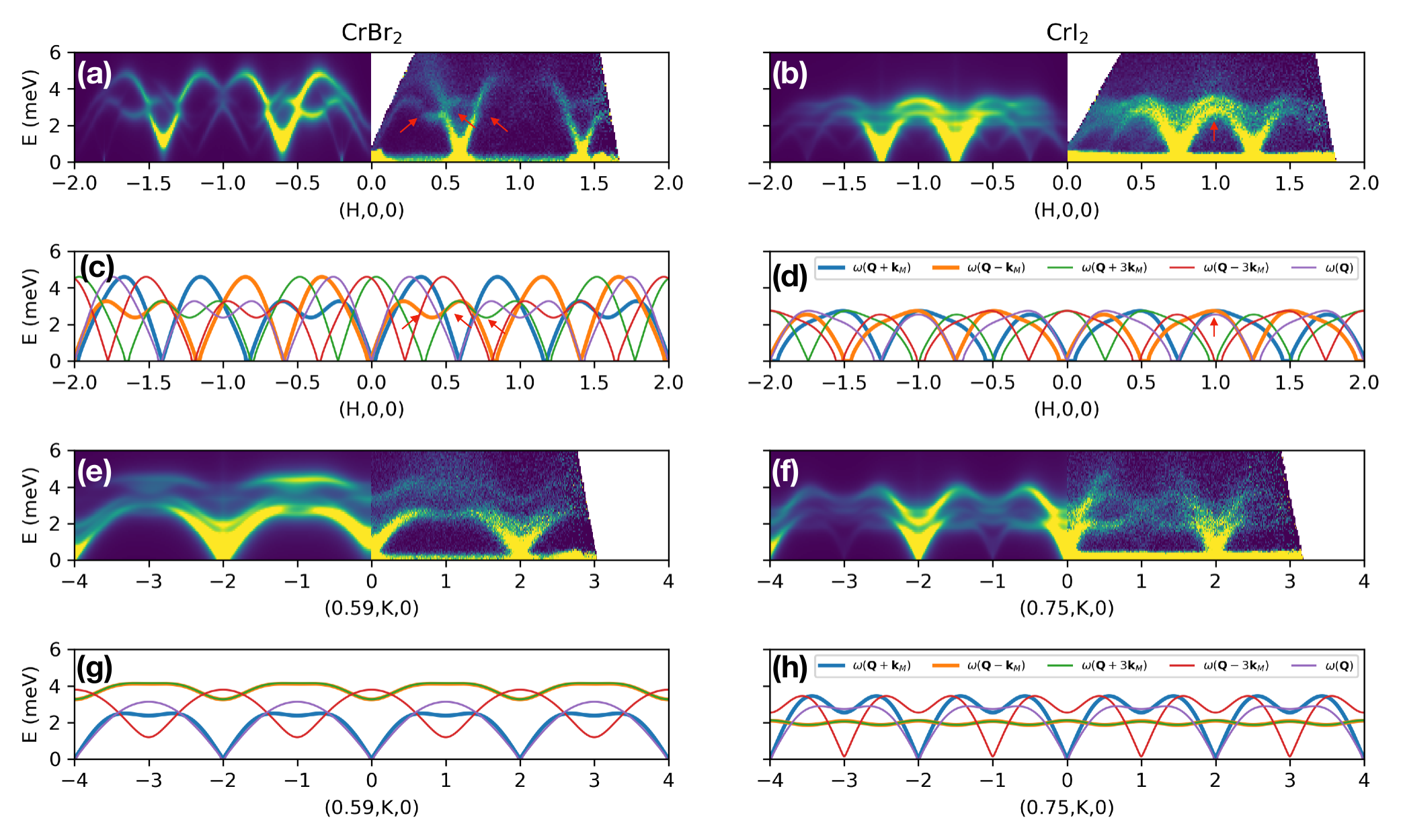}
\end{center}
\caption{Inelastic data at 5 K, similar to Fig.\ \ref{fig:Figure4}, but with the $\omega(\mathbf{Q} \pm 3\mathbf{k}_M)$ branches included in the simplified dispersion plots (c,d,g,h). Arrows indicate avoided crossings in the spin wave intensity along $(H,0,0)$ (a,b), and the corresponding points of intersection between spin-wave branches of the simplified dispersions.}
\label{fig:SuppInelasticBandCrossings}
\end{figure*}

In Fig.\ \ref{fig:SuppInelasticBandCrossings}, we show the inelastic data at 5 K along the ribbon-chain and interchain directions with simplified dispersions (omitting SIA and interlayer coupling), the same as for Fig.\ \ref{fig:Figure4} except that the dispersions include the $\omega(\mathbf{Q} \pm 3 \mathbf{k}_M)$ branches as well. (In Fig.\ \ref{fig:SuppInelasticBandCrossings}(a,b), we also include the $\omega(\mathbf{Q})$ branch omitted in Fig.\ \ref{fig:Figure4}.) Including the 3rd-harmonic branches allows us to fully account for the avoided crossings seen in the intensity along $(H,0,0)$. Arrows show the avoided crossings in Figs.\ \ref{fig:SuppInelasticBandCrossings}(a,b), and the corresponding points of interaction between branches for the simplified dispersion (without SIA or interlayer coupling)  in Figs.\ \ref{fig:SuppInelasticBandCrossings}(c,d). 

For CrBr$_2$, three avoided crossings are readily visible, corresponding to the following points of intersection: 
1) near $(0.33,0,0)$, a crossing between the primary $\omega(\mathbf{Q}-\mathbf{k}_M)$ branch and the 3rd-harmonic $\omega(\mathbf{Q}-3\mathbf{k}_M)$ branch; 2) near $(0.5,0,0)$, a crossing between the $\omega(\mathbf{Q}-\mathbf{k}_M)$, $\omega(\mathbf{Q}+\mathbf{k}_M)$, and $\omega(\mathbf{Q}+3\mathbf{k}_M)$ branches; and 3) near $(0.7,0,0)$. The last crossing is near points where all of the branches intersect, and it is difficult to tell which branches may be involved. For CrI$_2$ it is clear that the gap at $(1,0,0)$ corresponds to a crossing between the 
$\omega(\mathbf{Q}-\mathbf{k}_M)$ and 
$\omega(\mathbf{Q}+\mathbf{k}_M)$
branches. The inclusion of SIA creates openings at these intersection points, as seen by comparing the calculated intensity to the simplified dispersions.

\end{document}